\documentclass[twocolumn,prl,showpacs]{revtex4}
\usepackage{graphicx}
\usepackage{bm}
\begin{document}
\title{
Spin moment over 10-300 K and delocalization of magnetic electrons
above the Verwey transition in magnetite
}

\author{
Yinwan Li$^{1,2}$, P. A. Montano$^{1,3}$,
B. Barbiellini$^4$, P. E. Mijnarends$^{4,5}$,
S. Kaprzyk$^{4,6}$ and A. Bansil$^4$}

\affiliation{
$^1$Department of Physics, University of Illinois, Chicago IL 60680\\
$^2$Materials Science Division, Argonne National Laboratory, Argonne IL
60439\\
$^3$ Scientific User Facilities Division,
Basic Energy Sciences,
US Department of Energy,
1000 Independence Ave. SW,
Washington, DC 20585-1290\\
$^4$Physics Department, Northeastern University, Boston MA 02115 \\
$^5$Department of Radiation, Radionuclides \& Reactors,
Faculty of Applied Sciences,
Delft University of Technology, Delft, The Netherlands\\
$^6$Academy of Mining and Metallurgy AGH, 30059 Krak\'ow, Poland}

\date{\today}


\begin{abstract}
In order to probe the magnetic ground state, we have carried out
temperature dependent magnetic Compton scattering experiments on an
oriented single crystal of magnetite (Fe$_3$O$_4$), together with the
corresponding first-principles band theory computations to gain insight
into the measurements. An accurate value of the magnetic moment $\mu_S$
associated with unpaired spins is obtained directly
over the temperature range of 10-300K. $\mu_S$ is found to be non-integral
and to display an anomalous behavior with the direction of the external
magnetic field near the Verwey transition. These results reveal how the
magnetic properties enter the Verwey energy scale via spin-orbit coupling
and the geometrical frustration of the spinel structure, even though the
Curie temperature of magnetite is in excess of 800 K. The anisotropy of
the magnetic Compton profiles increases through the Verwey
temperature $T_v$ and indicates that magnetic electrons in the ground
state of magnetite become delocalized on Fe B-sites above $T_v$.
\end{abstract}

\maketitle
Magnetite was known to the ancient Greeks and its "magical"
properties have been a source of fascination for millenia. In
recent years, magnetite has become an archetype of complex
correlated materials which display competing orders and
interesting charge and orbital ordering effects in a solid state
setting. This complexity is accentuated by the expectation that
Fe in the inverse spinel crystal structure of magnetite will
assume a mixed valence. Simple chemical arguments suggest that on
the tetrahedral sites Fe occurs as ferric cations (Fe$^{3+}$),
while the octahedral sites are populated equally by ferric and
ferrous (Fe$^{2+}$) cations. Also, magnetite is important for
technological applications because it magnetizes spontaneously
like a ferromagnet but possesses low electrical conductivity so
that it can be used at high frequencies where eddy currents make
the use of iron impossible.

Magnetite has also drawn considerable attention because it
displays a first order phase transition--the Verwey
transition--at $T_v$=120K, below which the resistivity increases
by a factor of $\sim$100 \cite{verwey1,verwey2}. The earliest
interpretation was that this is an order-disorder transition
where the Fe$^{2+}$ and Fe$^{3+}$ cations on the octahedral sites
become ordered below $T_v$. However, this interpretation
and various pictures of electronic ordering
on different Fe sites still are quite hotly debated
issues \cite{novak,waltz,garcia,coey,radaelli}.
Band computations based on the
local density approximation (LDA) show that the bonding electrons
are shared in states that cannot be assigned solely to iron or
oxygen \cite{yanase0,zhang,yanase} and predict magnetite to be a
half-metal above $T_v$ where the minority spin electrons are
conducting, but the majority spins are insulating.

It is clear that the behavior of magnetic Fe 3$d$ electrons in magnetite
is the key to understanding the properties of this material. With this
motivation, we have carried out a magnetic inelastic x-ray scattering
investigation of an oriented single crystal of magnetite in the deeply
inelastic (Compton) regime. In this way we obtain a direct
measurement of the magnetic moment associated with unpaired spins in
magnetite over the wide temperature range of 10$-$300 K. Notably, magnetic
Compton scattering (MCS) is a genuine bulk probe which can detect the spin
magnetic moment ($\mu_S$) with a sensitivity of $\sim$0.1$\%$ (1 $\mu_B$
per 1000 electrons) \cite{cooper}. Previous estimates of $\mu_S$ are less
direct, being based on the application of sum rules for extracting orbital
moments ($\mu_O$) from x-ray magnetic circular dichroism (XMCD) spectra
\cite{huang}. Moreover, a recent XMCD study \cite{goering} of magnetite
reports a nearly vanishing $\mu_O$ and an integral value of $\mu_S$, which
is in sharp contrast to the earlier XMCD results \cite{huang} of a large
value of $\mu_O$ and a non-integral
$\mu_S$\cite{comment_goering,reply_huang}.
It is important therefore to determine independently whether $\mu_S$ is
integral or not and $\mu_O$ is small or large in order to understand the
role of spin-orbit coupling in magnetite.

An MCS experiment is well known to couple directly to the
wavefunction of magnetic electrons in the ground state of the
system \cite{cooper,li04,montano04}. The present MCS measurements
over the 10-300K range thus allow us to compare the
characteristics of the magnetic states in magnetite above and
below the Verwey transition. In particular, we show that the
anisotropy of the MCS spectrum becomes significantly larger as
the system goes through $T_v$. In order to help interpret our
measurements, we report extensive computations of the
magnetic momentum density within the first-principles band theory
framework. These and other model computations enable us to deduce
that the anisotropy of the MCS provides a handle for ascertaining
the localized vs delocalized nature of the magnetic electrons in
the ground state of magnetite.

In an MCS experiment one measures the magnetic Compton profile
(MCP), $J_{mag}(p_z)$, for momentum transfer along the scattering
vector $p_z$, which is defined by
\begin{equation}
J_{mag}(p_z) = J_{\uparrow}(p_z) - J_{\downarrow}(p_z)~,
\label{eq1}
\end{equation}
where $J_\uparrow$ ($J_\downarrow$) is the majority(minority)
spin Compton profile. $J_{mag}$ can be expressed in terms of a
double integral of the spin density, $\rho_{mag}({\mathbf p})$:
\begin{equation}
J_{mag}(p_z)= \int \int  \rho_{mag}({\mathbf p}) dp_x dp_y,
\label{eq2}
\end{equation}
where $\rho_{mag}({\mathbf p})\equiv \rho_{\uparrow}({\mathbf p})
- \rho_{\downarrow}({\mathbf p})$ is the difference of the
majority and minority spin densities. We emphasize that the
{\it area} under $J_{mag}(p_z)$ gives the spin moment
$\mu_S$. In contrast, the information about the nature of the ground state
wavefunction is described by the {\it shape} of
$J_{mag}(p_z)$.

Experiments were performed on beamline 11-B at the Advanced
Photon Source \cite{montano99} using an elliptical multipole
wiggler to generate circularly polarized photons. A high quality
single crystal of Fe$_3$O$_4$ was used and MCPs were measured
along the [100] and [110] crystal directions at temperatures of
10K, 100K, 140K, and 300K.
The Verwey transition temperature of the sample
was determined via resistivity measurements to be 120 $\pm$ 1 K.
The incident photon energy was 125 keV. The scattering angle
$\theta$ was 170$^{\circ}$. All measurements were made under an
external magnetic field of 7T oriented along the direction of the
scattering vector. The momentum resolution is estimated to be 0.4
a.u. (full-width-at-half-maximum). Magnetic calibration
measurements \cite{montano99} were performed using a crystal of
Fe oriented along [110].

The electronic structure was computed for the high temperature
(Fd3m) phase of Fe$_3$O$_4$ within an all-electron, fully charge
and spin self-consistent LDA-based band theory framework
\cite{bansil99}. All computations were carried out to a high
degree of accuracy, e.g., the crystal potential was converged to
better than $10^{-5}$ Ry. Our band structure is in good overall
accord with the few available studies
\cite{yanase0,zhang,yanase}. Various calculations agree on a
half-metallic ferromagnetic band structure with the Fermi level
crossing the minority spin $t_{2g}$ bands originating from the
octahedral Fe sites. The computed band structure and
wavefunctions were used to obtain spin-resolved momentum
densities $\rho_{\uparrow}({\mathbf p})$ and
$\rho_{\downarrow}({\mathbf p})$. These quantities were computed
over a fine mesh of $855 \times 10^6$ points within a sphere of
radius $\sim 13$ a.u. in ${\mathbf p}$ space \cite{mijnarends95}
in order to properly account for the fine structure in
$\rho_{\uparrow}$ and $\rho_{\downarrow}$, and provided the
dataset from which various projections of the spin-resolved
momentum density and the MCPs along the high symmetry directions
were computed.

\begin{figure} \begin{center}
\includegraphics[width=\hsize,width=9.cm]{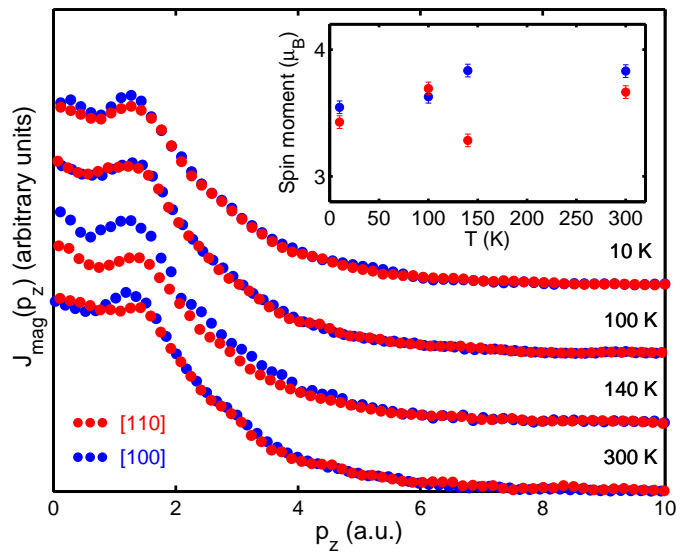}
\end{center}
\caption{(Color online)
$J_{mag}(p_z)$ for $p_z$ along [100] (blue dots) and [110] (red
dots) as a function of temperature. Different datasets are offset
vertically with respect to one another. Symbol size is
representative of error bars. Inset gives the magnetic moment of
unpaired spins (per formula unit) obtained from the area under
the MCP \cite{cooper}.
}
\label{fig1}
\end{figure}

The experimental spectra are summarized in Fig.~\ref{fig1}. The area under
the MCPs yields the magnetic moment $\mu_S$ associated with uncompensated
electronic spins shown in the inset. $\mu_S$ has a value of 3.54
$\mu_B$/formula unit at 10K for the magnetic field along [100]. The
uncertainty in $\mu_S$ is estimated to be $\pm 0.05$ $\mu_B$. By comparing
with the bulk magnetic moment of 4.05 $\mu_B$ \cite{aragon}, this yields a
corresponding orbital component of $0.51 \pm 0.05$ $\mu_B$. Interestingly,
the 140K dataset near the Verwey transition displays a striking anomaly in
that $\mu_S$ is 3.83 $\mu_B$ when the 7 T field is along [100], but for
the field along [110] it is 3.28 $\mu_B$. This directional preference of
$\mu_S$ weakens again at higher temperatures.

It is interesting to note that $\mu_S$ is non-integral. A simple
ionic picture gives $\mu_S=4\mu_B$ per formula unit \cite{waltz}.
In general, band theory can accommodate $\mbox{non-integral}$
$\mu_S$ since bands can be occupied partially. But, if spin is
assumed to be a good quantum number, it is straightforwardly
argued that $\mu_S$ in a half-metallic system must also be
strictly integral. However, a non-integral $\mu_S$ value can be
obtained if the spin-polarized band calculation is supplemented
by spin-orbit interactions.
The spin-orbit coupling has been estimated to be of the
order of 10 meV \cite{mcqueeney},
so that we would expect a substantial
orbital moment due to Fe$^{2+}$.

Bulk magnetization measurements in magnetite \cite{abe,matsui} reveal a
significant magneto-crystalline anisotropy. This anisotropy is relatively
small above $T_v$ with [111] being the easy axis, while below $T_v$, the
anisotropy is larger with [001] being the easy axis. However, the maximum
saturation field is about 2T, so that under the field of 7T used in the
present measurements, we would not normally expect anisotropic effects.
Therefore, changes in the spin moment with the direction of the magnetic
field seen in the inset to Fig.~\ref{fig1},
which are especially striking at 140K,
are anomalous and do not fit within the standard picture. The origin of
this anomaly is unclear, but it may be related to the effect of
geometrical frustration on magnetic properties above $T_v$
\cite{radaelli}.
%
%
Notably, Ref.~[\onlinecite{uzu}] considers a Hamiltonian for the 
pyrochlore B lattice with a spin-orbit interaction. In this model, several 
complex orbital ordered states with noncollinear orbital moments are found 
as a possible ground state for the high-temperature phase. This degeneracy 
of the lowest energy states can be broken by an external magnetic field in 
a manner which depends on the field direction. Since the energy 
differences between these states are very small, a magnetic field of 7T 
can be sufficient to switch between the states \cite{private_uzu}. Further 
MCS measurements of the spin moment in magnetite over a range of 
temperatures around $T_v$ should prove worthwhile in this connection. Here 
a note should also be made of the role of magnons, which will reduce the 
moment below the canonical value of $4$ $\mu_B$ at high temperatures 
\cite{neel}.

\begin{figure}
\begin{center}
\includegraphics[width=\hsize,width=9cm]{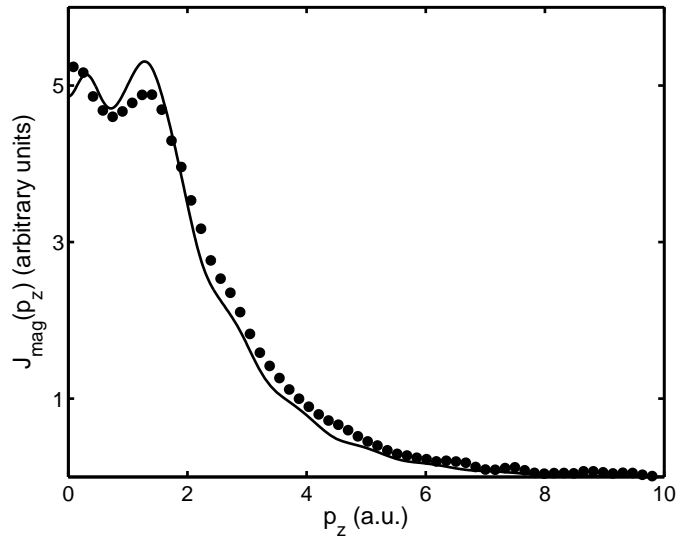}
\end{center}
\caption{
Shape of the 140K [110] experimental MCP of Fig.~\ref{fig1} is
compared with corresponding theoretical (resolution broadened)
MCP obtained within the conventional LDA-based band theory
framework.
}
\label{fig2}
\end{figure}

Some insight into the shape of the MCPs can be obtained through
the LDA-based first-principles band theory computations. This is
illustrated in Fig.~\ref{fig2}, which shows a good level of
agreement between the shapes of the computed and measured [110]
MCPs. The LDA correctly reproduces the overall width of the
measured MCP as well as the pronounced peak at 1.3 a.u. The
shallow bumps in the experimental MCP at momenta between $2-3$
a.u. and $4-5$ a.u. are similar to features in the theoretical
MCP. This fine structure is the hallmark of
interference terms that cannot be explained within a purely
atomic picture.

%
%

In order to extract wavefunction localization features, 
we now discuss the
anisotropy of the MCP defined as the difference between the two
directional MCPs \cite{footnote_anis1},
\begin{equation}
\Delta J_{mag} =\left\langle J_{mag}^{[100]} \right\rangle 
- J_{mag}^{[110]},
\label{eq3}
\end{equation}
where the
brackets $\left\langle \ldots\right\rangle$ denote an average
over temperature. 
$\left\langle J_{mag}^{[100]} \right\rangle$ is normalized 
at each temperature to the integral of the corresponding 
$J_{mag}^{[110]}$, so that
$\Delta J_{mag}$ integrates to zero.
The average $\left\langle J_{mag}^{[100]} \right\rangle$
provides a temperature-independent "background" for 
highlighting changes in the shape of $J_{mag}^{[110]}$ as a 
function of temperature.\cite{footnote_anis2}

\begin{figure}
\begin{center}
\includegraphics[width=\hsize,width=9.cm]{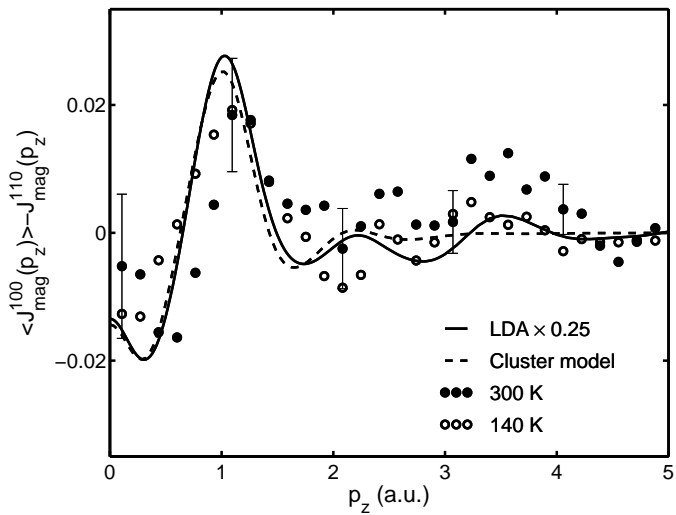}
\end{center}
\caption{
Anisotropy of the MCP defined as 
$\Delta J_{mag} =\left\langle J_{mag}^{[100]} \right\rangle 
- J_{mag}^{[110]}$, where the areas under all
profiles are normalized to one. Experimental results for the 140K and 300K
datasets are compared with the anisotropy obtained within the
conventional LDA-based band theory framework and the model
cluster computation discussed in the text. Note, the LDA curve
has been scaled down by a factor of 4.
}
\label{fig3}
\end{figure}

Fig.~\ref{fig3}, which focuses on the 140K and 300K data above the 
Verwey transition, shows that although the LDA reproduces the overall features in
the measured anisotropy, the amplitude of the anisotropy given by the LDA
is too large by about a factor of 4 (note that plotted LDA curve is scaled
down). In order to gain insight into the shape and amplitude of $\Delta
J_{mag}$, we have considered this quantity for a variety of clusters of Fe A-
and B-site atoms and their nearest neighbor (NN) Fe atoms, using linear
combinations of Slater type orbitals (STOs) where the cluster wavefunction
is constructed by allowing STOs on NN Fe atoms to mix via a small
admixture parameter $f$ \cite{chiba}.  Our results show that the
shape of the experimental $\Delta J_{mag}$, including the location of the
peak at 1.12 a.u., can only be reproduced for a B-site Fe cluster in which
the NN Fe atoms are situated at a NN distance $d_{BB}$=$5.61$ a.u. ($2.97$
\mbox{\AA}) along the [110] direction \cite{footnote_D}.
In this case, the momentum
wavefunction obtained by Fourier transforming the cluster wavefunction
contains an oscillating factor of $[1+2f\cos\{d_{BB}(p_x+p_y)/\sqrt{2}\}]$,
which yields the correct shape, and as Fig.~\ref{fig3} shows, for $f=0.1$
the computed curve (dashed) gives a good agreement in shape and
amplitude with the 140K and 300K measurements in the low momentum region. 
We note that $\Delta J_{mag}$ is rather insensitive to the inclusion of covalency
with O atoms, and furthermore, the lack of oscillations in the cluster
model at $p_z > 2$ a.u. is related to inaccuracy in our STOs near the
nucleus and is of little interest in the present context of solid state
effects.

\begin{figure}
\begin{center}
\includegraphics[width=\hsize,width=9.cm]{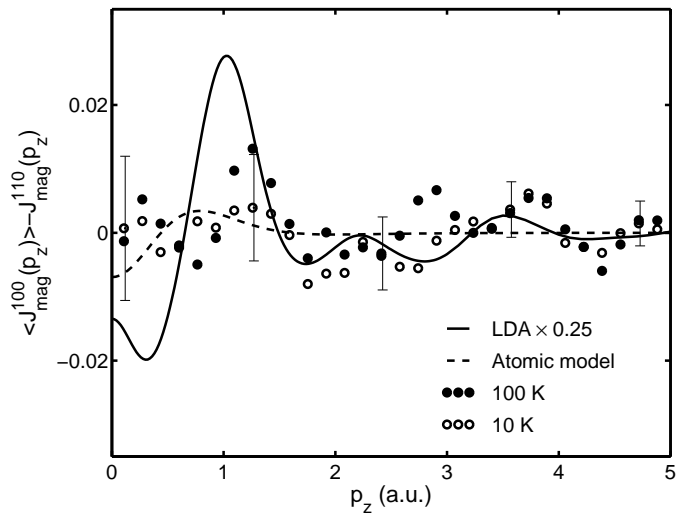}
\end{center}
\caption{
Anisotropy of the MCP defined as 
$\Delta J_{mag} =\left\langle J_{mag}^{[100]} \right\rangle 
- J_{mag}^{[110]}$,
where the areas under all profiles are normalized to
one. Experimental results for the 10K and 100K datasets are compared
with the simple atomic analysis of Ref.~\cite{montano04}.
The LDA curve is reproduced from Fig.~\ref{fig3} for reference.
}
\label{fig4}
\end{figure}

In order to delineate the temperature dependence of $\Delta J_{mag}$ the 
results for the 10K and 100K datasets below $T_v$ are presented in 
Fig.~\ref{fig4}. The peak located at $1.12$ a.u. is seen now to have 
essentially disappeared and the anisotropy is strongly reduced compared to 
even the scaled LDA curve. The rather flat result of Fig.~\ref{fig4} is 
consistent with a value of the admixture parameter $f \approx 0$ 
\cite{footnote_E} and with the simple atomic analysis of 
Ref.~\cite{montano04}. A similar result for $f$ was obtained from an 
analysis of a positron annihilation study on magnetite carried out below 
$T_v$ \cite{chiba}.

To be more quantitative we define the anisotropy amplitude
\begin{equation}
A = \frac{1}{2} \int_0^{p_{max}}
\left|  \Delta J_{mag}(p)  \right| dp,
\label{eq4} 
\end{equation}
where $p_{max}=5$ a.u. is a momentum cutoff \cite{footnote_A}. As we have 
already discussed in connection with Figs.~\ref{fig3} and~\ref{fig4}
above, the strength of  the peak at $1.12$ a.u. in $\Delta J_{mag}(p)$ 
is related to the extent to which 
magnetic electrons on B-sites develop phase coherence via the admixture 
parameter $f$ on NN Fe B-sites. It may then be argued that the quantity 
$A$ defined via the integral of Eq.~\ref{eq4} is a measure of the area 
under the peak in $\Delta J_{mag}(p)$, and thus of the number of magnetic electrons 
participating in this coherence process \cite{footnote_B}.

\begin{figure}
\begin{center}
\includegraphics[width=\hsize,width=9.0cm]{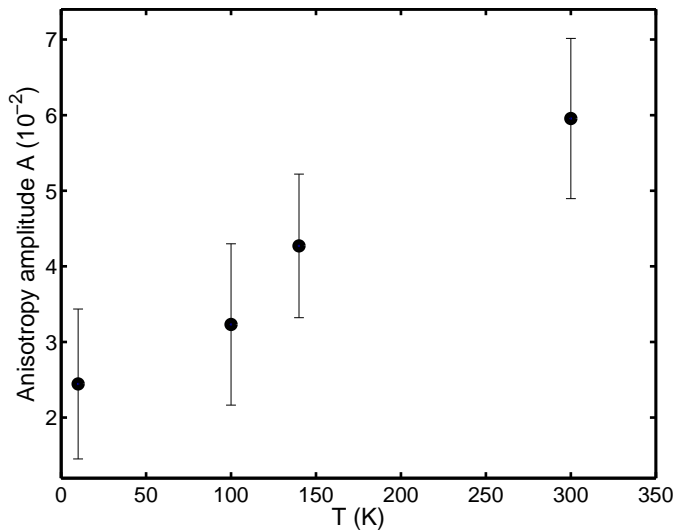}
\end{center}
\caption{
Amplitude $A$ of the MCP anisotropy obtained from Eqs. 3 and 4 as a 
function of temperature. 
}
\label{fig5}
\end{figure}

Fig.~\ref{fig5} shows that the value of $A$ increases 
systematically in going from the 10K to the 300K dataset. Interestingly, 
this behavior of $A$ correlates with that of the admixture parameter $f$ 
discussed above in connection with Figs.~\ref{fig3} and~\ref{fig4} in 
that, like $A$, the parameter $f$ also increases with temperature. In 
other words, the Verwey transition is accompanied by a change in the 
character of the ground state wavefunction such that the wavefunction 
becomes significantly more delocalized on Fe B-sites above $T_v$. 
Notably, the present analysis suggests that the 
B-site is involved in the change in valence across the Verwey transition
as in Ref.~\cite{nazarenko}, 
and not the A-site as argued in Ref.~\cite{rozenberg} recently.
Positron annihilation experiments \cite{positrons,biasini}, which also 
give information on spin-resolved momentum density, have revealed 
covalency effects between Fe and O atoms but have failed to detect 
significant changes at $T_v$. This result however may reflect the tendency 
of the positron to sample mostly O-ions and interstitial spaces rather 
than the Fe-ions. 

In conclusion, we have presented magnetic Compton scattering measurements 
on a magnetite single crystal, together with corresponding predictions of 
the MCP within the conventional band picture. An accurate value of the 
unpaired spin moment $\mu_S$ is thus obtained directly over a wide 
temperature range. For example, at 10K $\mu_S$ has clearly a non integer 
value of $3.54 \pm 0.05$ $\mu_B$/formula unit for the magnetic field along 
[100] demonstrating a non vanishing spin-orbit coupling. The ground state 
of magnetite is shown to be remarkably sensitive to the direction of the 
external magnetic field around the Verwey transition and to display an 
anomalous magnetic moment, which may be a manifestation of a large 
degeneracy and associated geometrical frustration in the spinel lattice. 
The amplitude of the anisotropy of the MCP is shown to increase with 
temperature. We argue on this basis that the ground state wavefunction in 
the system becomes delocalized on Fe B-sites above $T_v$.

\acknowledgments

We acknowledge discussions with R.S. Markiewicz, Hsin Lin and E. Tosatti,
the support of the U.S.D.O.E. contracts DE-AC03-76SF00098,  
DE-FG02-07ER46352 and the U.S. DOE
support of the Advanced Photon Source at Argonne National Laboratory. We
benefited from the allocation of supercomputer time at NERSC, Northeastern
University's Advanced Scientific Computation Center (ASCC) and the
Stichting NCF (Foundation National Computer Facilities).


\begin{thebibliography}{99}
\bibitem{footnote_anis2}
We have also obtained $\Delta J_{mag}$ where the average 
$\left\langle J_{mag}^{[100]} \right\rangle$ 
was taken over just the two high (140K and 300K) or the two 
low (10K and 100K) temperature datasets. The resulting 
anisotropy is quite similar to that shown in Figs. 3 and 4 
where all four datasets are employed 
in the averaging process, indicating that our use of 
$\left\langle J_{mag}^{[100]} \right\rangle$ 
as background does not introduce any spurious 
features in the anisotropy defined by Eq. 3.
\bibitem{footnote_D}
The other Fe-Fe distances
$d_{BA}=6.58$ a.u. and $d_{AA}=6.88$ a.u. do not yield the main peak
at $1.12$ a.u. Only the distance $d_{BB}=5.61$ a.u.
yields the good fit shown in Fig.~\ref{fig3}.
\bibitem{footnote_E}
The error in our values of $f$ is estimated to be $\pm 0.02$.
\bibitem{footnote_A}
The temperature dependence of $A$ shown in Fig.~\ref{fig5}
is not sensitive to the value of the cut-off momentum $p_{max}$.
\bibitem{footnote_B}
The factor of $1/2$ in
Eq.~\ref{eq4} accounts for the presence of
positive excursions in $\Delta J_{mag}(p)$
(in particular the peak at $1.12$ a.u.)
which involve the same area as the negative
contributions.
\end{thebibliography}
\end{document}